\documentclass[
    ,final            
]
  {aipproc}
  \usepackage{amssymb,amsmath,graphicx}

\layoutstyle{8x11double}

\newcommand{\be}{\begin{equation}}
\newcommand{\ee}{\end{equation}}

\def\bY{{\bf Y}}
\def\bU{{\bf U}}
\def\bV{{\bf V}}
\def\bmm{{\bf m}}
\def\tl{{\tilde{L}}}
\def\bmM{{\bf M}}
\def\la{\lambda}
\def\ga{\gamma}
\def\unity{{\hbox{1\kern-.8mm l}}}
\def\BR{{\rm BR}}

\begin{document}

\title{Lepton flavour violation in the supersymmetric type-II seesaw mechanism}


\author{F.~R.~Joaquim}{
  address={Departamento de F\'isica Te\'orica \& Instituto de F\'isica
  Te\'orica UAM/CSIC, Facultad de Ciencias C-XI, Universidad
  Aut\'onoma de Madrid, Cantoblanco, E-28049 Madrid, Spain}
}

\author{Anna Rossi}{
  address={Dipartimento di Fisica ``G.~Galilei'', Universit\`a di
Padova I-35131 Padua, Italy} }

\begin{abstract}
We summarize the predictions for the radiative decays $\ell_j
\rightarrow \ell_i \gamma$ within the context of the supersymmetric
type II seesaw mechanism considering universal boundary conditions
for the soft SUSY breaking terms. The dependence on the low-energy
neutrino parameters is discussed and the deviations from the
analytical results for large $\tan\beta$ analyzed.
\end{abstract}

\maketitle


The observation of lepton flavour violation (LFV) signals (besides
those manifest in neutrino oscillations) would be a clear and
dramatic manifestation of {\it new physics} since, in the SM, they
are strongly suppressed by the smallness of neutrino masses. A
typical example where alternative LFV processes can be sizeable is
the minimal supersymmetric standard model (MSSM) where they can be
enhanced through one-loop exchange of sleptons (and gauginos) if
their masses are not too far from the electroweak scale and do not
conserve flavour.

Regarding flavour violation, most of the available studies rely on
the most conservative scenario of universal sfermion masses at a
high scale $\Lambda$. In such cases, flavour non-conservation in
the sfermion masses might arise from renormalization group (RG)
effects in the presence of flavour-violating Yukawa
couplings~\cite{lfv1}. In the so-called type I seesaw
mechanism~\cite{seesaw}, these couplings are of the form $\bY_N N
L H_2$ where $L$ is a lepton doublet, $H_2$ the hypercharge +1/2
Higgs doublet of the MSSM and $N$ a heavy neutrino singlet which
decouples at a scale $M_N$. From $\Lambda$ to $M_N$ the presence
of $\bY_N$ induces LFV in the sfermion masses. On the other hand,
the effective neutrino mass matrix after electroweak symmetry
breaking reads $\bmm_\nu=v_2^2\,\bY_N^T\,\bmM_N^{-1}\,\bY_N$ where
$\bmM_N$ is the heavy neutrino mass matrix. Therefore, one would
expect to find a connection between low-energy neutrino parameters
and the amount of flavour-violation induced in the slepton sector.
This turns out to be impossible since one cannot reconstruct the
high-energy neutrino parameters ($\bY_N$ and $\bmM_N$) from
low-energy measurements in the type I seesaw framework. One is
therefore compelled to consider model-dependent conditions in
order to make some predictions~\cite{Raidal:2008jk}.

Let us now discuss what happens in the type II seesaw
mechanism~\cite{ss2} where the MSSM particle is extended by adding
a vector-like pair of heavy triplets which transform under the
$SU(2)_W\times U(1)_Y$ gauge group as $T\sim (3,1), \bar{T} \sim
(3,-1)$~\cite{hms,ar}. Besides the usual MSSM interactions one can
write the following new terms in the superpotential\footnote{One
can also include the term $\frac{1}{\sqrt{2}}\la_1 H_1 T H_1$ but
this will be irrelevant for our discussion.}
\be\label{L-T}%
W_T=\frac{1}{\sqrt{2}}\bY^{ij}_{T} L_i T L_j + \frac{1}{\sqrt{2}}
\la H_2 \bar{T} H_2 +  M_T T \bar{T}\,,
\ee%
where $i,j = e, \mu, \tau$ are family indices. $\bY^{ij}_{T}$ is a
$3 \times 3$ symmetric matrix, $\la$ is a dimensionless
unflavoured coupling and $M_T$ denotes the mass parameters of the
triplets.

At the electroweak scale the Majorana neutrino mass matrix emerges
and is given by
\be
\label{T-mass}%
{\bf m}^{ij}_\nu =\frac{v_2^2 \la}{M_T} \bY^{i j}_T\;\;\;,\;\;\;
\bmm_\nu = {\bf U}^\star {\rm diag}(m_1, m_2, m_3) {\bf
U}^\dagger\,, \ee
where $m_i$ are the effective neutrino masses and $\bU$ is the
low-energy leptonic mixing matrix which can be written as
\be%
\label{Udef}%
{\bf U}={\bf V}\cdot{\rm diag}({1,\rm e}^{ i\phi_1},
{\rm e}^ { i\phi_2})\;\;,\;\; {\bf V}\equiv {\bf
V}(\theta_{12},\theta_{23},\theta_{13},\delta)\,.
\ee %
We have denoted the three mixing angles by $\theta_{12}$,
$\theta_{23}$ and $\theta_{13}$, and the ``Dirac'' and
``Majorana'' CP-violating phases by $\delta$ and $\phi_{1,2}$,
respectively. The matrix ${\bf V}\equiv {\bf
V}(\theta_{12},\theta_{23},\theta_{13},\delta)$ can be
parameterized as the CKM mixing matrix in terms of $\theta_{ij}$
and $\delta$. The neutrino masses can be expressed as:
\begin{eqnarray}
\label{mdef}%
\!\!\!\!\!\!\!\!{\rm NO}&\!\!\!\!\!\!\!:&
\!\!\!\!\!\!m_2^2=m_1^2+\Delta
m^2_\odot\;\;,\;\;m_3^2=m_1^2+\Delta
m^2_{\rm atm}\nonumber \\
\!\!\!\!\!\!\!\!{\rm IO}&\!\!\!\!\!\!:&
\!\!\!\!\!\!m_2^2=m_3^2+\Delta m^2_{\rm atm}+\Delta
m^2_{\odot}\,\,,\,\,m_1^2=m_3^2+\Delta m^2_{\rm atm}\,,
\end{eqnarray}
where NO (IO) stands for considering a normal (inverted) order for
the neutrino mass spectrum. The current allowed ranges for the
mixing angles and neutrino mass-squared differences may be found
in Ref.~\cite{GonzalezGarcia:2007ib}.

The presence of the Yukawa interactions $\bY_T$ may also induce LFV
in the spleton mass matrix $\bmm_\tl$. This was first pointed out by
Rossi in Ref.~\cite{ar} in a framework where the soft sfermion
masses are universal at a scale above $M_T$, \emph{i.e.}
$\bmm_\tl=m_0^2\unity$.

Between the universality scale $\Lambda$ and $M_T$ the RG induced
LFV matrix elements read~\cite{ar}:
\begin{equation}
\label{m2lind} %
(\bmm^2_{\tl})_{ij}\simeq - \frac{9 \,m_0^2 + 3 \,a_0^2}{8 \pi^2}
\left(\, \bY_{T}^\dag
\bY_{T}^{}\,\right)_{ij}\ln\frac{\Lambda}{M_T}\;,(i\neq j).
\end{equation}
The most striking feature of this result is that the combination
$\bY_{T}^\dag \bY_{T}^{}$ can be traded\footnote{Note that this is
true only when the RG effects on the flavour structure of
$\bY_{T}^\dag \bY_{T}^{}$ are negligible. See discussion at the
end.} by $\bmm^{\dagger}_\nu \bmm_\nu^{}$ using Eq.~(\ref{T-mass})
and therefore the dependence of $(\bmm^2_{\tl})_{ij}$ on the
low-energy neutrino parameters is now direct, contrarily to what
happens in the type I seesaw mechanism. In particular, and taking
into account Eqs.~(\ref{T-mass})-(\ref{mdef})
\begin{eqnarray}
\label{m2lU}%
{\rm NO}&\!\!\!\!\!\!\!:& \!\!\!\!\!\!(\bmm^{\dagger}_\nu \bmm_\nu
)_{ij}=m_1^2\,\delta_{ij}+ \Delta m^2_\odot
\bV_{\!i2}^{}\bV_{\!j2}^{\ast} + \Delta m^2_{\rm atm}
\bV_{\!i3}^{}\bV_{\!j3}^{\ast}\,,\nonumber \\
{\rm IO}&\!\!\!\!\!\!:& \!\!\!\!\!\! (\bmm^{\dagger}_\nu \bmm_\nu
)_{ij}=m_3^2\,\delta_{ij}+ (\Delta m^2_{\rm atm}+
 \Delta m^2_\odot)
\bV_{\!i2}^{}\bV_{\!j2}^{\ast} \nonumber\\
&&\quad\quad\quad\quad+\Delta m^2_{\rm atm}
\bV_{\!i1}^{}\bV_{\!j1}^{\ast} \,,
\end{eqnarray}
which reveals the fact that the LFV induced in the slepton masses
does not depend on the lightest neutrino mass ($m_1$ and $m_3$ for
the normal and inverted mass spectrum,
respectively)~\cite{Joaquim:2006uz}. The existence of LFV in the
slepton sector may lead to the enhancement of certain LFV
processes which otherwise are very suppressed. Typical examples
are the radiative decays $\ell_j \rightarrow \ell_i \gamma$ which
can be naively estimated by
\be
\label{BRdef} {\rm BR}(\ell_j \to \ell_i +\gamma) \simeq
\frac{\alpha^3}{G^2_F}
\frac{|(\bmm^2_{\tilde{L}})_{ji}|^2}{m_S^8}\tan^2\beta~ {\rm
BR}(\ell_j \to \ell_i \nu_j\bar{\nu}_i)\,. \ee
Here, $\alpha$ and $G_F$ denote the fine structure and Fermi
constant, respectively, and $m_S$ a typical SUSY mass for the
sparticles inside the loops. It is convenient to work with the
quantities~\cite{ar}
\begin{eqnarray}
&&\!\!\!\!\!\!\!\!\!\!\!\!\!\!\!\!\!\!\!\!\!\!\!\!R_{\tau\mu}\equiv\frac{\BR(\tau
\to \mu\ga)}{\BR(\mu \to e\ga)} \simeq \left|\frac{( \bmm^{2
}_{\tilde{L}})_{\tau \mu}}{( \bmm^{2 }_{\tilde{L}})_{\mu
e}}\right|^2 \frac{\BR(\tau \to \mu \nu_\tau \bar{\nu}_\mu)}
{\BR(\mu \to e \nu_\mu \bar{\nu}_e)}\,,
\nonumber \\
\label{Rs} &&\!\!\!\!\!\!\!\!\!\!\!\!\!\!\!\!\!\!\!\!\!\!\!\!R_{\tau
e}\equiv\frac{\BR(\tau \to e\ga)}{\BR(\mu \to e\ga)} \simeq
\left|\frac{( \bmm^{2 }_{\tilde{L}})_{\tau e}}{( \bmm^{2
}_{\tilde{L}})_{\mu e}}\right|^2 \frac{\BR(\tau \to e \nu_\tau
\bar{\nu}_e)} {\BR(\mu \to e \nu_\mu \bar{\nu}_e)}\,,
\end{eqnarray}
which do not depend (in most cases) on $\tan\beta$ and on the SUSY
spectrum. In the above equations $\BR(\tau \to \mu \nu_\tau
\bar{\nu}_\mu)/\BR(\mu \to e \nu_\mu \bar{\nu}_e)=0.1737$ and
$\BR(\tau \to e \nu_\tau \bar{\nu}_e)/\BR(\mu \to e \nu_\mu
\bar{\nu}_e)=0.1784$.
\begin{figure}
\label{Fig1}%
\begin{tabular}{c}
  \includegraphics[height=.24\textheight]{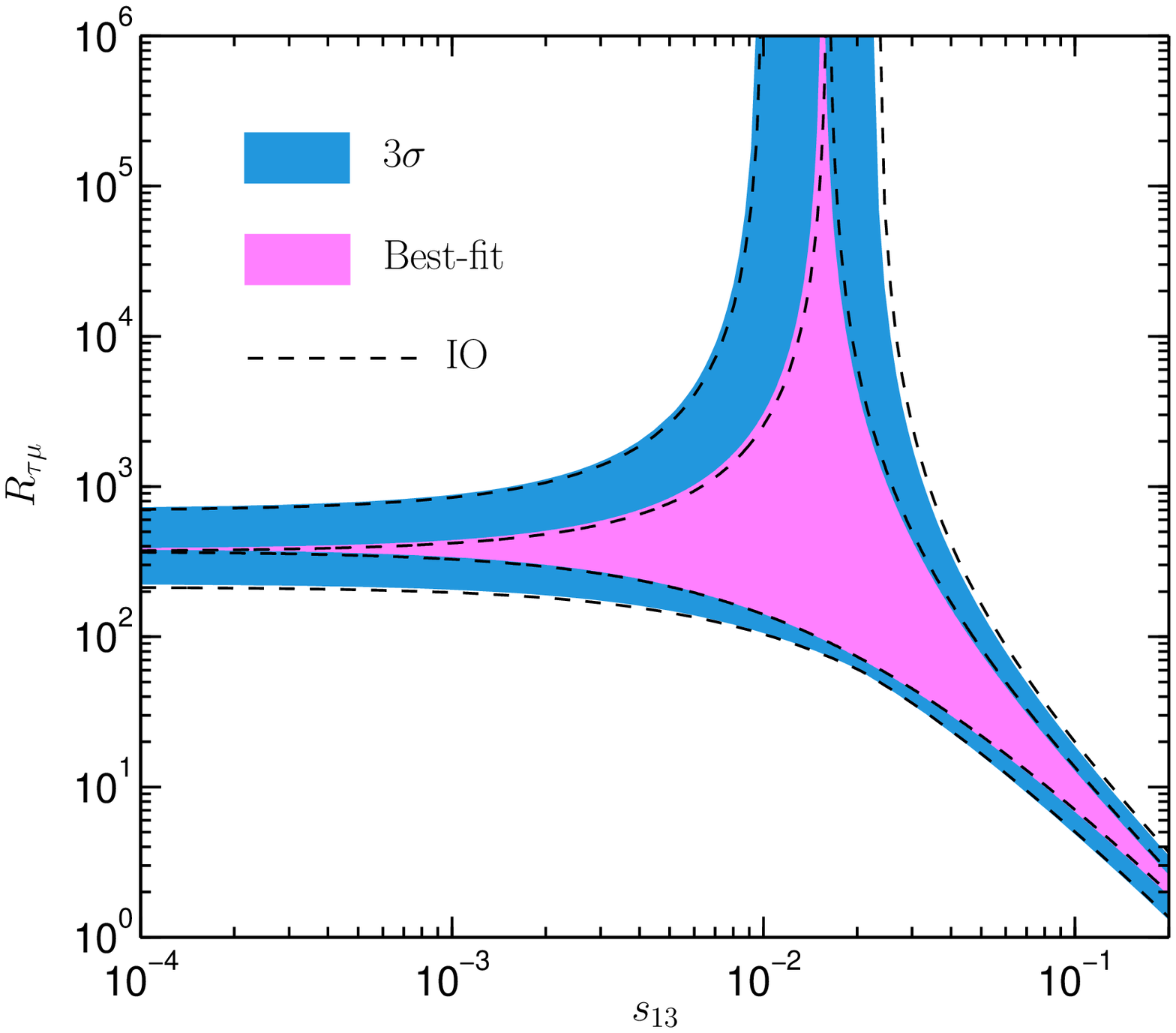} \\
  \includegraphics[height=.24\textheight]{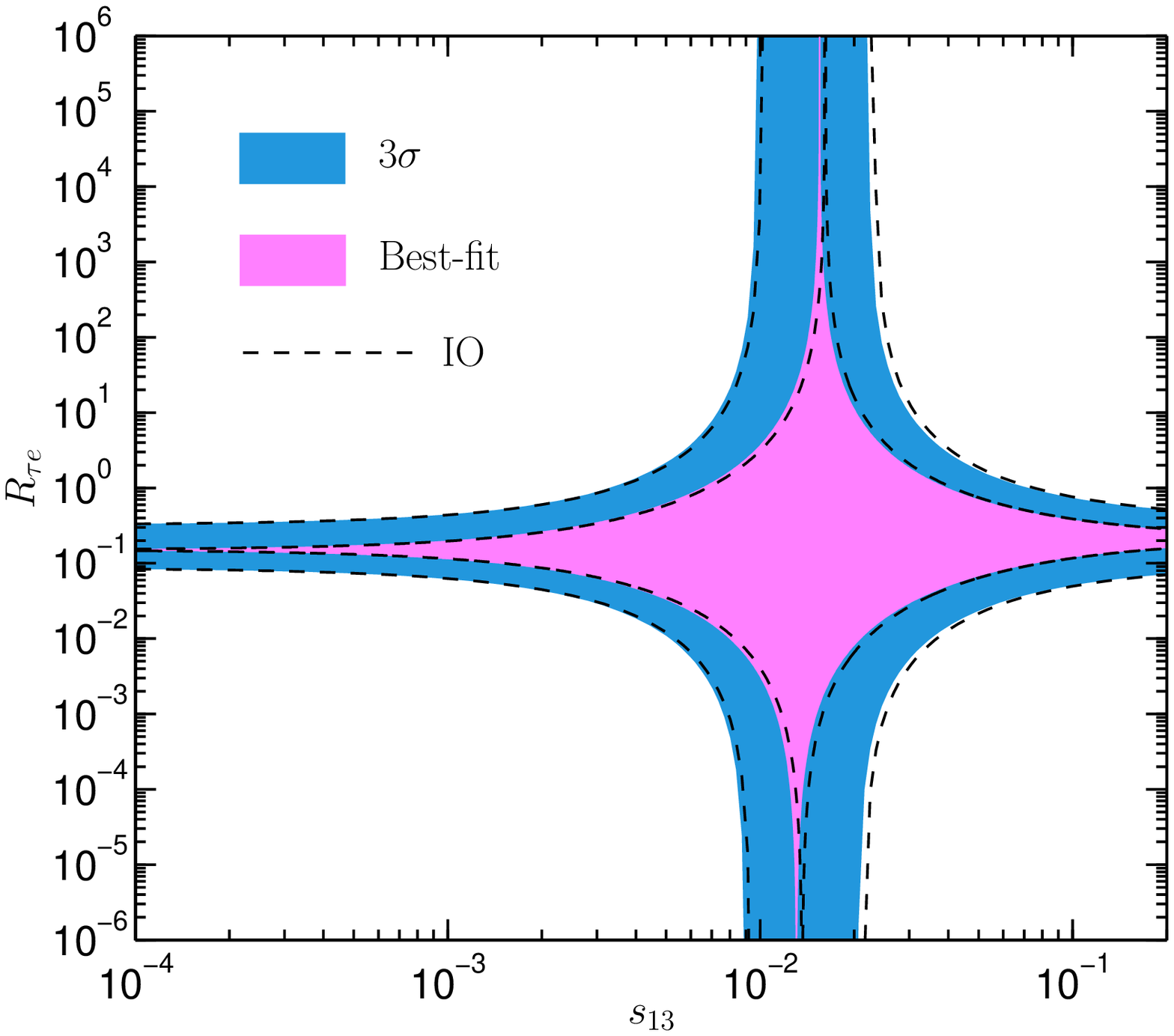}
\end{tabular}
\caption{Allowed ranges for $R_{\tau\mu}$ (upper plot) and $R_{\tau
e}$ (lower plot) as a function of $s_{13}$. The shaded regions
correspond to the NO case and the dashed lines delimit the same
regions for the IO spectrum.}
\end{figure}

Using now a standard parameterization for the elements of $\bV$ we
can write {\small
\begin{eqnarray}
\left|\left[\bmm^{2 }_{\tilde{L}}\right]_{\mu
e}\right|^2&\!\!\!\!\!\propto& \!\!\!\!\!\frac{c_{13}^2}{4}
\left[\,\rho^2\,c_{23}^2\sin^2(2\theta_{12})+4\,s_{13}^2\,
s_{23}^2\right.\nonumber\\%
&&\left.\pm
2\,\rho\,s_{13}\,\cos\delta\,\sin(2\theta_{12})\sin(2\theta_{23})\right]\,,
\nonumber \\
\left|\left[\bmm^{2 }_{\tilde{L}}\right]_{\tau
e}\right|^2&\!\!\!\!\!\propto& \!\!\!\!\!\frac{c_{13}^2}{4}
\left[\,\rho^2\,s_{23}^2\sin^2(2\theta_{12})+4\,s_{13}^2\,
c_{23}^2\right.\nonumber\\%
&&\left.\mp
2\,\rho\,s_{13}\,\cos\delta\,\sin(2\theta_{12})\sin(2\theta_{23})\right]\,,
\nonumber\\
\label{m2ltaumu}\left|\left[ \bmm^{2 }_{\tilde{L}}\right]_{\tau
\mu}\right|^2&\!\!\!\!\!\propto& \!\!\!\!\!\frac{1}{4} \left\{\,\,
\rho^2s_{13}^2\,\sin^2\delta\sin^2(2\theta_{12})\right.\nonumber\\
&&\!\!\!\!\!\!\!\!\!\!\!\!\!\!\!\!\!\!\!\!\!\!\!\!\!\!\!\!\!\!\!\!\!\!\!
\!\!\!\!\!\!\!\!\!\!\!\!\!\!\!\!\!\!\!\!\!\!\!\!\!\!\!\!\!\!\!\!
\left.+\left[\,c_{13}^2\,\sin(2\theta_{23})\mp\rho\,s_{13}\,
\cos\delta\,\sin(2\theta_{12})\cos(2\theta_{23})\,\right]^2\right\}\,.
\end{eqnarray}}%
and use this to express $R_{\tau\mu}$ and $R_{\tau e}$ in terms of
the low-energy neutrino parameters. We have used the notation
$s_{ij}\equiv \sin\theta_{ij}$, $c_{ij}\equiv \cos\theta_{ij}$ and
$\rho \equiv \Delta m^2_{\odot}/\Delta m^2_{\rm atm}$. In the
limit $s_{13}\to 0$ one gets $R_{\tau\mu}\simeq
0.7\,s_{23}^2/[\rho^2\,\sin^2(2\theta_{12})]\simeq 383.0$ and
$R_{\tau e}\simeq0.18\tan^2\theta_{23}\simeq 0.2$. For
$s_{13}\simeq 0.2$ one obtains $R_{\tau\mu}\simeq 0.17
c_{23}^2\,\cot^2\theta_{13} \simeq 2.0 $ and $R_{\tau
e}\simeq0.18\cot^2\theta_{23}\simeq 0.2$. In Fig.~\ref{Fig1} we
present the allowed ranges for $R_{\tau\mu}$ and $R_{\tau e}$
considering the experimental allowed ranges for the neutrino
parameters and varying the phase $\delta$ between 0 and $2\pi$.
The strong enhancements and suppressions observed in the plots for
$|s_{13}|\simeq \rho\cot\theta_{23}\sin(2\theta_{12})/2\simeq
0.016$ are due to cancelations in $|[\bmm^{2 }_{\tilde{L}}]_{\mu
e}|^2$ and/or $|[\bmm^{2 }_{\tilde{L}}]_{\tau e}|^2$ for
$\delta=0,\pi$ [see Eqs.(\ref{m2ltaumu}) and
Refs.~\cite{Joaquim:2006uz}].
\begin{figure}
\label{Fig2}%
\begin{tabular}{c}
  \includegraphics[height=.23\textheight]{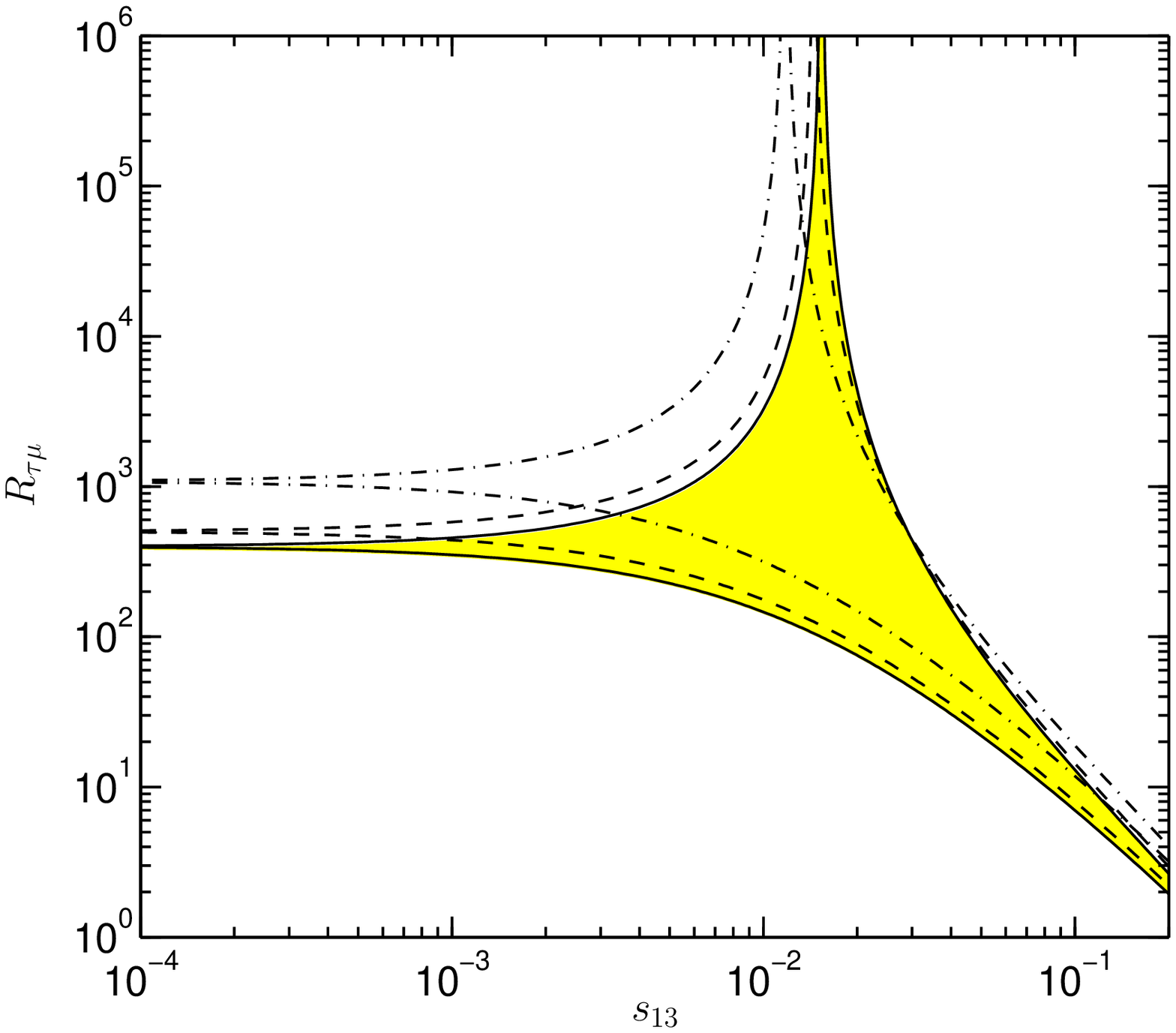} \\
  \includegraphics[height=.23\textheight]{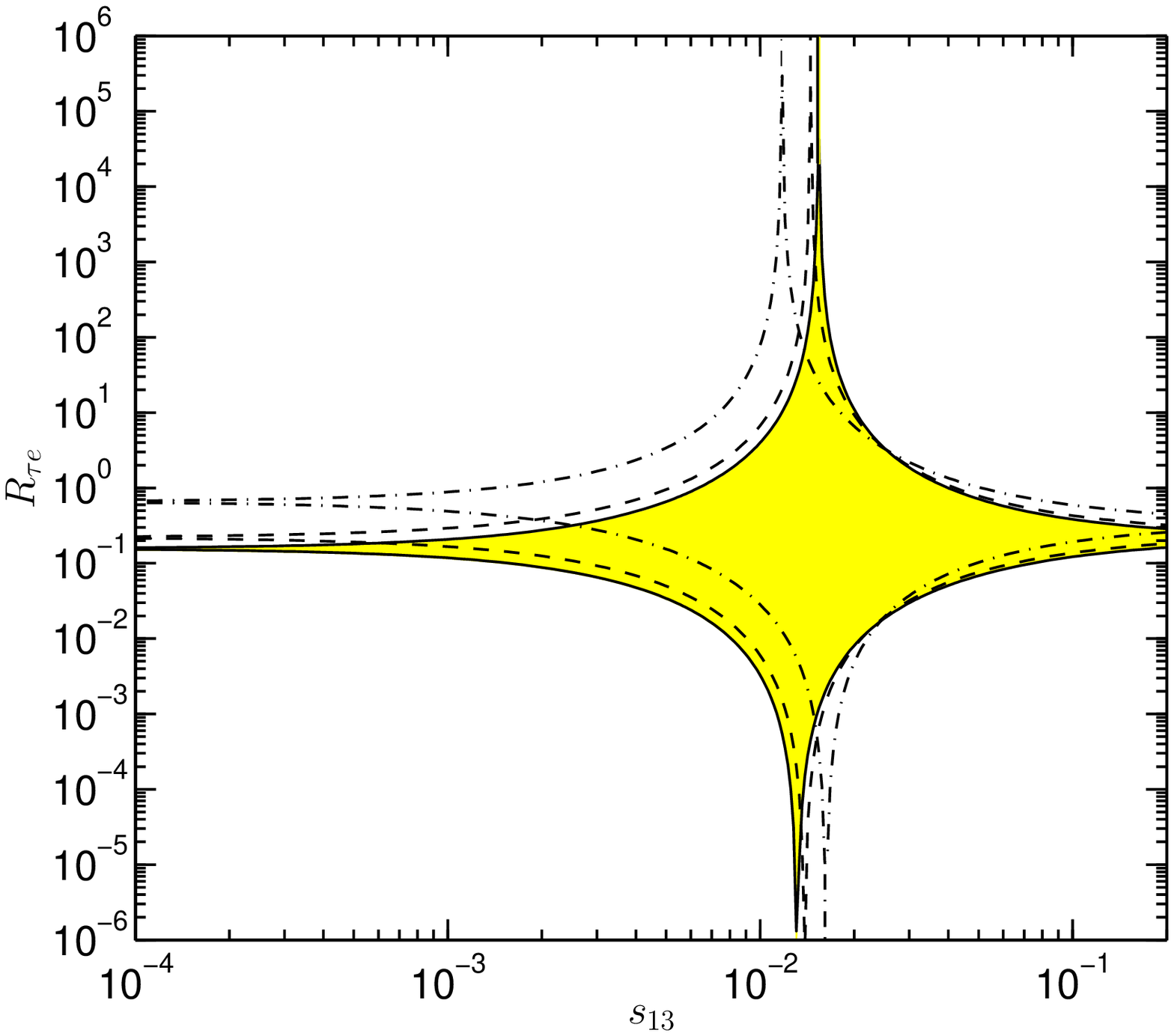}
\end{tabular}
  \caption{$R_{\tau\mu}$ (upper plot) and $R_{\tau
e}$ (lower plot) as a function of $s_{13}$. The shaded regions
correspond to the analytical results using the best-fit values for
the neutrino parameters and varying $\delta$ from 0 to $2\pi$. The
lines delimit the same region using the full numerical calculation
of the processes and performing the running of all couplings and
masses from low to high-energies and back. The solid, dashed and
dash-dotted lines correspond to $\tan\beta=10,30,50$, respectively.}
\end{figure}
The above results have been obtained using the estimate given in
Eq.~\ref{BRdef}. However, this estimate is valid in the limit of
equal masses for the sparticles entering in the loop. Also notice
that the above results have been obtained using the low-energy
neutrino parameters. This approximation is valid if the RG effects
in $\bY_T$ are negligible when one goes from low to high energies.
This might not be the case if $\tan\beta$ is large. Moreover, for
large $\tan\beta$, the left-handed staus get separated in mass
from the smuon and selectron. This is due to running effects in
the diagonal elements of the slepton mass matrix when the Yukawa
of the $\tau$ is large. Putting all together, we expect that the
results for $R_{\tau \mu}$ and $R_{\tau e}$ deviate from the ones
presented above when $\tan\beta$ is large. In Fig.~\ref{Fig2} we
present the results taking into account the best-fit values for
the neutrino parameters. We see that as $\tan\beta$ increases, the
results deviate from the ones given by the analytical expressions
(which are shown in shaded). For instance, for $\tan\beta=50$  and
small $s_{13}$, $R_{\tau \mu}$ is enhanced by a factor of
approximately 2.5 with respect to the analytical result. This will
be discussed in more detail in a forthcoming
publication~\cite{prep}. It is clear from Figs.~\ref{Fig1} and
\ref{Fig2} that if $s_{13}$ is close to the present upper bound
(let's say $s_{13}\simeq 0.2$) then the present bound ${\rm
BR}(\mu\to e \gamma)< 1.2 \times 10^{-11}$ implies ${\rm
BR}(\tau\to \mu \gamma) \lesssim 10^{-10}$ and ${\rm BR}(\tau\to e
\gamma) \lesssim 10^{-11}$, which are both beyond future
experiments. Therefore, if $s_{13}$ is large, the only decay which
could be observed is $\mu\to e \gamma$.




\begin{theacknowledgments}
F.~R.~J. is supported by the EU 6th Framework Program
MRTN-CT-2004-503369 ``The Quest for Unification: Theory Confronts
Experiment''.
\end{theacknowledgments}



\bibliographystyle{aipproc}   

\bibliography{sample}


\end{document}